\documentstyle[epsfig]{article}
\begin{document}

\title{Kolmogorov-Smirnov test as a tool to study \\
 the distribution of ultra-high energy cosmic ray sources}

\author{D. Harari, S. Mollerach and E. Roulet\\ 
CONICET, Centro At\'omico Bariloche,\\
Av. Bustillo 9500, Bariloche, 8400, Argentina}
\maketitle
\begin{abstract}
We analyze in detail the two-dimensional Kolmogorov-Smirnov test as a tool to
learn about the distribution of the sources of the ultra-high energy cosmic
rays. We confront in particular models based on AGN observed in X rays, on
galaxies observed in HI and isotropic distributions, discussing how this
method can be used not only to reject isotropy but also to support or reject
specific source models, extending results
obtained recently in the literature.  
\end{abstract}

\section{Introduction}

One of the most puzzling aspects of cosmic rays (CRs) is the enormous energies
they can reach, which can be up to seven orders of magnitude higher than those
being achieved in the most powerful human-made accelerator.
The astrophysical sites where the acceleration of these particles takes place 
have then
to be quite extreme, and their identification is of paramount importance to
understand the CR origin and the mechanisms capable of producing them.

For observed energies larger than $E_{GZK}= 6\times 10^{19}$~eV,  CRs cannot
come from very far away because if they were produced e.g. from beyond $\sim
200$~Mpc, no matter what was the initial value of their energy at the source
it would likely be degraded below $E_{GZK}$ by the time they reach us, due to the
interactions with the cosmic microwave background that take place 
during the CR propagation \cite{gzk}. 
  A suppression of the CR flux at the
highest energies compatible with this phenomenom has indeed been measured 
recently \cite{hiresspec,augerspec}. This suggests
that above $E_{GZK}$ only nearby sources contribute to the observed CR
fluxes. Moreover, since at these high energies the CR trajectories are not
expected to suffer very large deflections in the galactic or extragalactic
magnetic fields (at least if  the CR charges are not too large), it is to be
expected that the inhomogeneous distribution of the potential sources in our
cosmic neighborhood  will give rise to an anisotropic pattern of CR arrival
directions.

The search for the sources of the ultra-high energy (UHE) CRs got renewed
interest  after data from the Pierre Auger Observatory revealed
 a correlation between the arrival
directions of the highest energy events (with $E>57$~EeV, where 1~${\rm
  EeV}\equiv 10^{18}$~eV) and the directions towards the active galactic
nuclei (AGN) closer than about 100~Mpc \cite{agn}. 
The maximum significance for this
correlation was obtained for angular separations between events and AGN
smaller than $3.2^\circ$ and for AGN closer than 71~Mpc, for which 20 events
out of 27 were found to correlate while only 5.6 correlations were expected to
arise by chance from an isotropic distribution. This indeed suggests that the
angular deflections are not too large at the highest energies (although of
course the closest AGN to an event need not necessarily be its source) and
gives further support to the idea that the spectral suppression is due to the
GZK effect, not just to the exhaustion of the acceleration power of the 
sources.
The correlation found in data from the Auger Observatory
by no means proves that the AGN are the actual 
sources of the CRs, because these active galaxies could well be acting as 
tracers of the nearby large scale structure which also hosts other types of
galaxies or possible acceleration sites (such as gamma ray bursts). 
Even if AGN are the sources, it may
well be that only a particular subclass of the vast compilation contained in
the V\'eron-Cetty and V\'eron catalog (VC catalog)
 used to establish the correlation,
or even some obscured AGN absent in the catalog, are the actual sources of
the highest energy CRs. 
Hence, further analyses can be important to shed some light on this issue.

There have been in particular two recent analyses trying to find indications
about the possible UHECR source population using the two dimensional
Kolmogorov-Smirnov  (2DKS) test.
The first by George et al. \cite{ge08}
 confronted the arrival directions of the Auger
events above 57~EeV with the positions of the X ray selected AGN in the SWIFT
BAT catalog \cite{bat}, updated with the first 22 months of data. 
The second, by Ghiselini et al. \cite{gh08}, confronted those same events with the
HIPASS catalog  of HI selected radio galaxies from the All-Sky Parkes Survey
 \cite{me04,wo06}. 
The first analysis concluded that  although the standard 2DKS didn't give much
hints of correlations, restricting the AGN to those closer than 100~Mpc and 
 weighting them by their X ray luminosities led to a
correlation of $\sim 98$\%, giving hence additional 
support to the Auger Observatory results. 
The second analysis, weighting the HIPASS galaxies  by their HI flux found a
correlation of only $\sim 72$\%, which increased to 99\% after restricting the
test to the southern equatorial 
sample (which is more complete) and to the more massive
galaxies (with HI mass greater than $1.1\times 10^{10}\ M_\odot$). 
In this work we want to reconsider this kind of analyses and discuss further
issues which can help to understand what is being tested with the 2DKS
method,  hoping that this will be useful for future studies of this type.

\section{The 2DKS test}

In one dimension $X$, the Kolmogorov-Smirnov test looks for the maximum value $D$ of
 the difference between the fractional 
cumulative distribution of the variable $X$ measured in the data and that
 expected for 
a model. The significance of that departure  can be obtained analytically from
 the expected distribution of $D$ in the case of the null-hypothesis (data
 drawn from the model) or alternatively 
 estimated from the  fraction of samples  simulated
according to the model which 
 give rise to larger departures $D$  than the data themselves. 
Note that the maximum departure will always be found at one of the values of
 $X$ realized by the data, as can be easily understood from the step-like
 behavior of the cumulative distribution.
Also note that by looking to the cumulative distribution at one of the data
 points $X_i$, the one-dimensional KS test compares
 the fraction of data points with 
$X< X_i$  with the corresponding fraction expected in the model.

The generalization of this method for two dimensions, $X$ and $Y$, consists of
looking for the fractions in each of the four natural quadrants (adopting a
specific set of coordinates, typically the equatorial ones for astronomical
applications) of both the data and a model, which may correspond e.g. to
directions drawn from the locations of the
objects contained in a given catalog. 
Then one finds the reference point and the quadrant
for which the difference between the two fractions is maximal. In the
original proposal of Peacock \cite{peacock}, 
the reference points are all the possible
combinations $X_i,\ Y_j$, where $i$ and $j$ refer to any of the data
points. In the modified test proposed by Fasano and Franceschini \cite{ff},
 which is much less
demanding computationally and has similar power, one just chooses as reference
points the locations of the different data points $X_i,\ Y_i$. 
Alternatively, in the so-called two-sample test one first finds the distance
$D_1$ obtained using all data points as reference, then the distance $D_2$
obtained using instead the coordinates of the objects in the catalog as
reference points, and finally considers their 
average $\overline D\equiv(D_1+D_2)/2$
as a measure of the distance between the two distributions. For large data
(and catalog) sets the two approaches should give similar results.
 For definiteness we will consider hereafter  the distances obtained in the
two-sample case.

What the 2DKS test ultimately probes is whether the data are distributed in
the sky in the same proportion as the model, and achieves that by checking if
there is any quadrant in which the fraction of the data points is
significantly different from the corresponding 
fraction of the catalog objects present. In
particular, large deviations are expected if the data points are unrelated to
the model considered, but also if they correspond to a subsample of the
objects in the model having a different overall distribution. Similarly, if
only a subset of the catalog objects 
considered are the actual sources, the extra ones
may ruin the agreement in the comparison, and only after appropriate cuts the
source population may be identified using this method.
 On the other
hand, if the actual sources are different from those in the model but have
anyway a similar spatial distribution (e.g. one being a tracer of the other) 
the 2DKS 
test will indicate a good agreement even if the model does not correspond to
the true sources. George et al.\cite{ge08}
 generalized this method by considering the
fractions obtained after weighting each object in the catalog
 by the relative exposure of the
Observatory in the corresponding region of the sky and also weighting
it by its 
flux. Since ref.~\cite{ge08} considered as potential sources the AGN
identified by SWIFT, they were weighted by the flux measured in 
X rays, as could be expected if the CR luminosities were proportional to the X
ray ones. Similarly ref.~\cite{gh08} obtained the fractions for the HIPASS
based model weighting the galaxies by their integrated HI fluxes as well as 
 the exposure of the Auger Observatory.  

A probability of correlation was then quoted as a measure of the
significance of the results. It is defined as the fraction of isotropic
simulated sets of arrival directions giving departures $D$ larger 
than those found in the actual
data, comparing both the real and simulated sets  to the same source 
model, i.e. to the same reference catalog.
 Note that this gives a measure of the discrepancy
between the data and the isotropic simulations rather than a direct measure of
the agreement between the data and the model, i.e. it tells how worse the
isotropic simulations do under a similar test rather than how typical of
a model realization the data are. 
One of the purposes of the present paper is
to introduce both ingredients into the application of the 2DKS test to the
search for a satisfactory UHECR source model.
To stress the point that the correlation probability gives just a measure of
the departure from isotropy we will 
hereafter refer to it instead by the name of `anisotropy probability'. 

Let us finally note that if the data are a fair realization of the model being
tested, the distances $D$ obtained will just be the result of statistical 
fluctuations due to the limited sampling,  and are hence expected to decrease
with the size of the data sample $n$ as $n^{-1/2}$.
% (in the limit in which the
%number of catalog points is very large or that an exact expression for the
%model expectations is available). 
It is then customary to
use as an alternative to $D$ the quantity $Z=D\sqrt{n}$, which would remain
approximately constant with increasing $n$. However, if the data are not a
fair sample of the model there will be a typical distance between the true
distribution  sampled by the data and the model being
considered. This means that there will be one particular quadrant with respect
to one particular sky direction for which the difference in the fractions of
the true distribution and the model distribution is maximal, and this
difference $D$ will have a certain fixed value to which the data will
ultimately approach with very large statistics. Hence,
 in this case the quantity
$Z$ is expected to grow with the size of the data sample. 

When the two-sample case mentioned previously is considered, an effective value
$n_{eff}\equiv n_1 n_2/(n_1+n_2)$ is adopted in the definition $Z=\overline D
\sqrt{n_{eff}}$, with $n_1$ the number of data points and $n_2$ the number of
catalog points \cite{peacock}.
Anyway, since the significance is obtained
 by comparing the data with the results
of isotropic simulations with the same number of events, it doesn't matter
whether $\overline D$ or $Z$ are compared. 
We will hence display the results for
$\overline D$ (omitting hereafter the overline) 
because the interpretation in terms of  the difference in fractions
has a  more direct meaning than the quantity $Z$.
We note that when comparing the expected distances for a  model compared
against itself, an analytic expression for the probability to obtain a certain
value can be derived \cite{ff,singh}, 
but in the general case in which the model is different
that the reference catalog the use of simulations is required.

\section{Results}

\begin{figure}[ht]
\centerline{{\epsfig{width=2.in,file=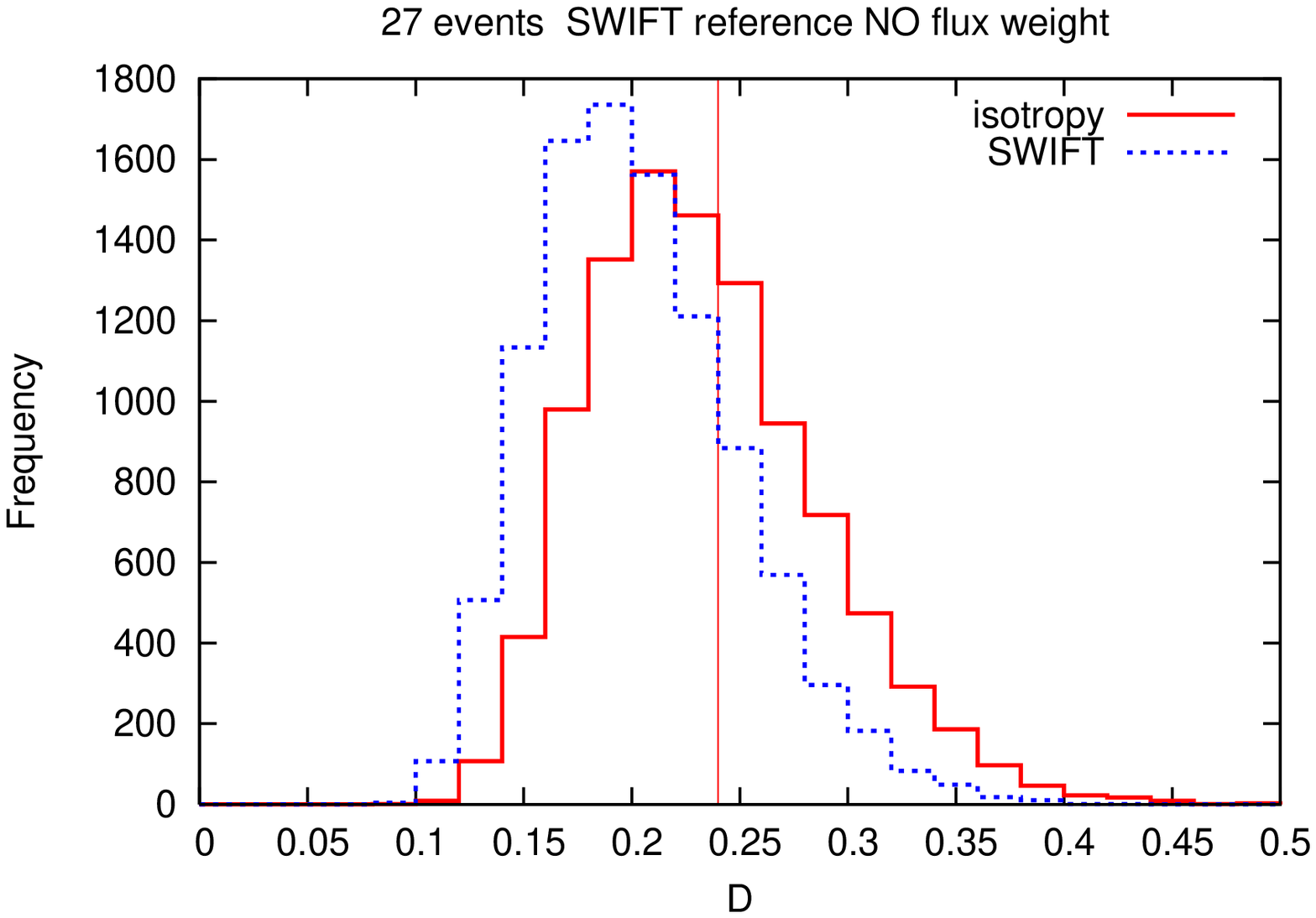,angle=-90}\epsfig{width=2.in,file=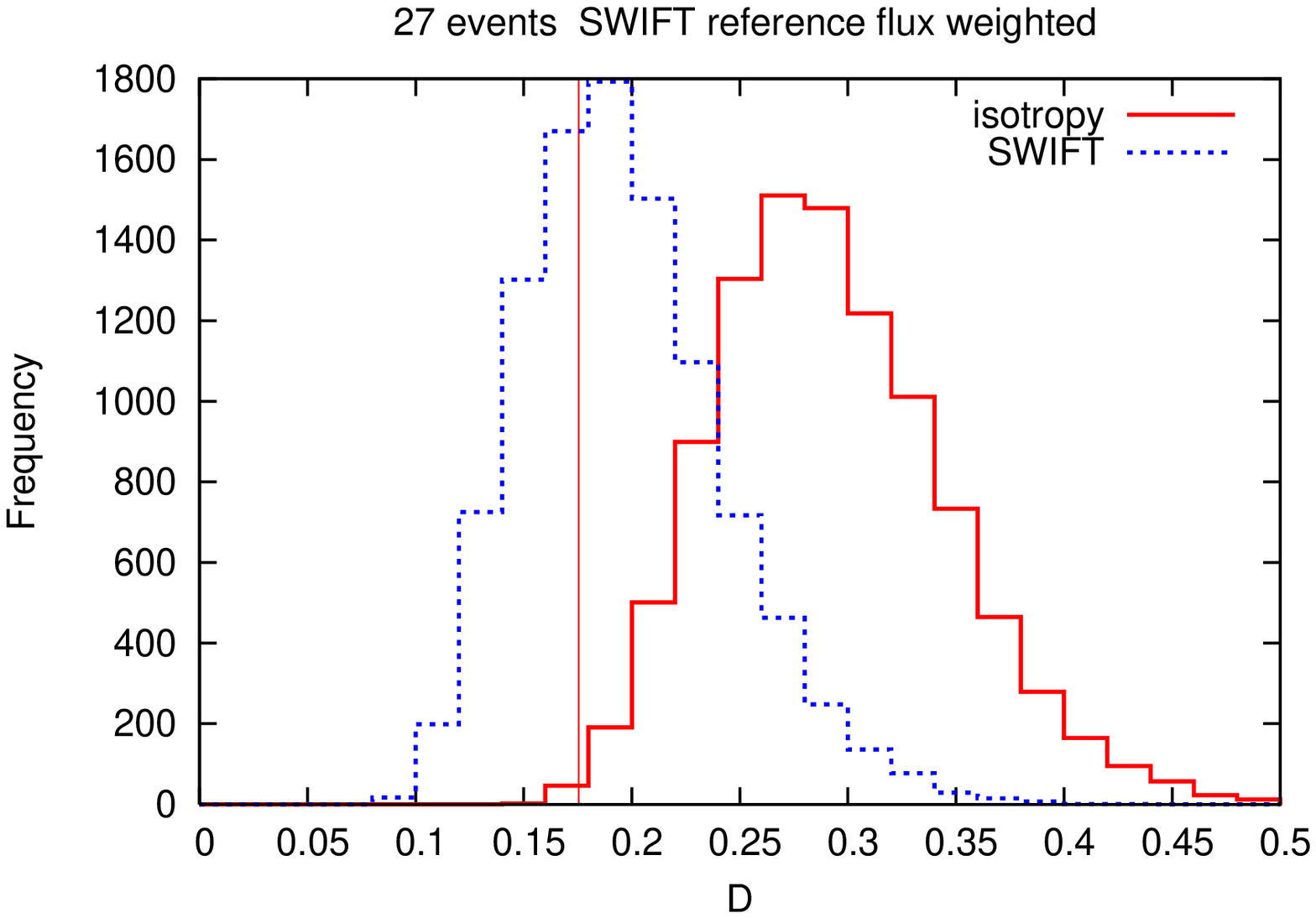,angle=-90}}}
\centerline{{\epsfig{width=2.in,file=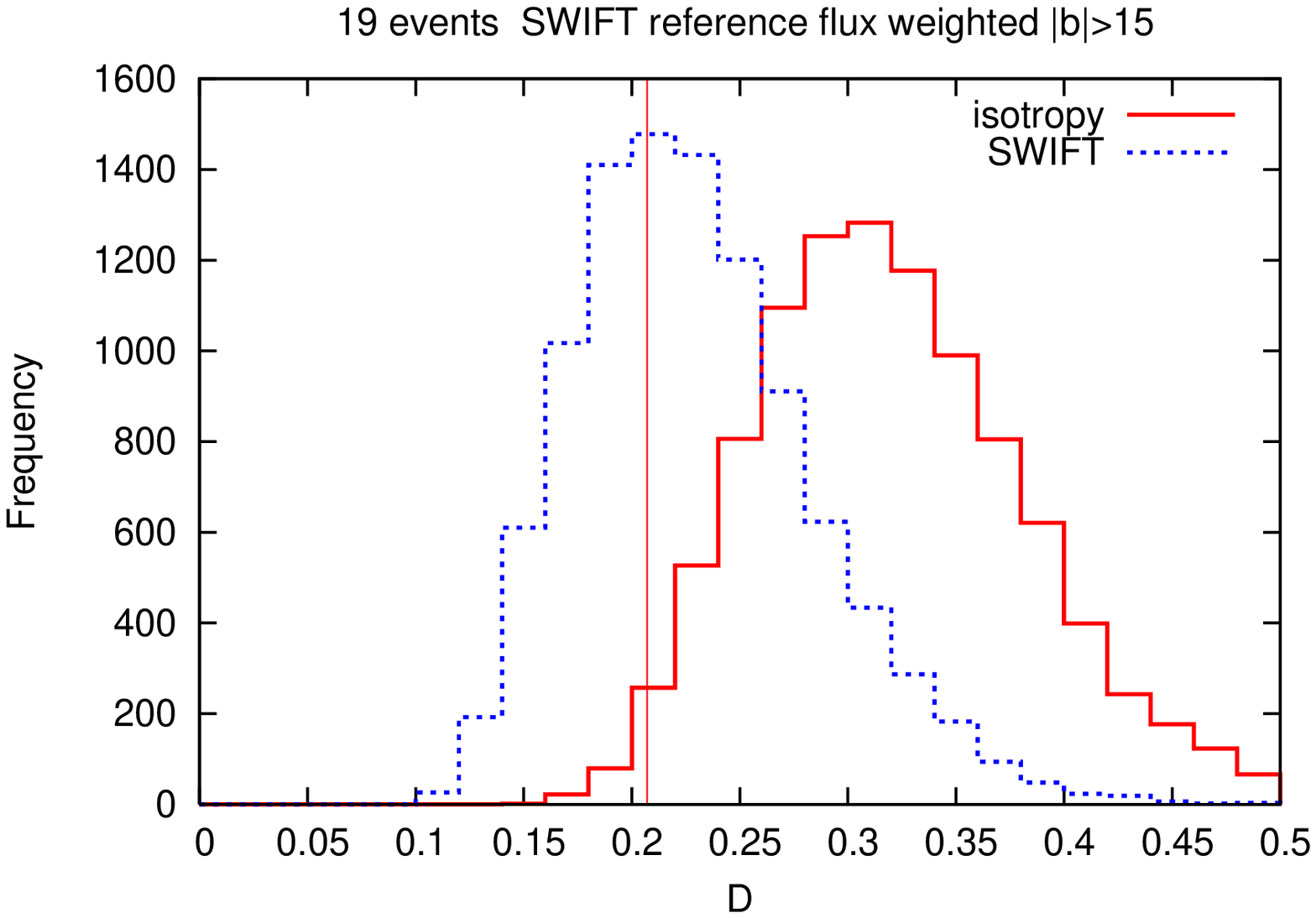,angle=-90}\epsfig{width=2.in,file=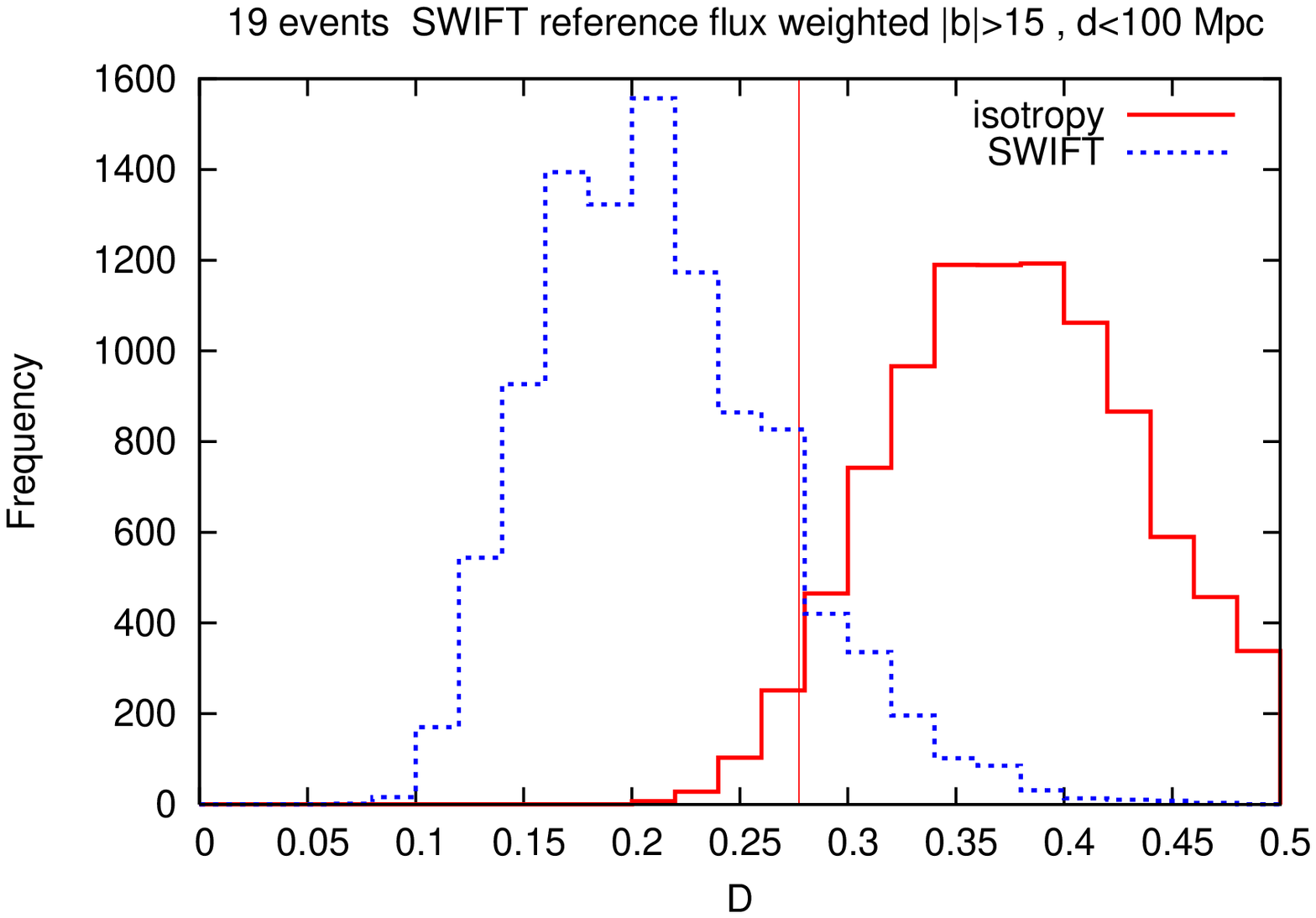,angle=-90}}}
\caption{Distribution of 2DKS distances $D$  between the weighted fractions
  obtained in different AGN models and the fractions in simulated data samples
  (according to the same AGN models or to isotropy). The vertical lines
  correspond to the values obtained for the Auger data.} 
 \label{fig14.fig}
\end{figure}

Fig.~1 shows the resulting distances taking as reference objects  
the AGN in the 22 months SWIFT BAT catalog \cite{bat22}
 with different cuts and weights. This catalog contains 163 AGN identified in
 X rays with reliably measured flux and redshift and within the field of view
 of the Auger Observatory  (corresponding to declinations $\delta<24.8^\circ$ 
 for the showers with zenith angle smaller than $60^\circ$).
In the  top-left panel all these sources  are included,  weighting them
by the relative exposure of the experiment. We always use the
equatorial coordinate system to identify the four quadrants with respect to
the different reference directions (we have checked that adopting a different
coordinate system, such as the galactic one, does not change the qualitative
behavior of the results obtained).
The two histograms  shown correspond to the
values obtained for sets of 27 events simulated  according to an isotropic 
distribution  or to one generated following the catalog.  The vertical
line is the distance obtained with the 27 Auger events \cite{agn}. 
It is seen that there
is an almost complete overlap between the two histograms, what prevents the
possibility  of discriminating  clearly between the two scenarios (SWIFT
AGN sources or isotropy) using this method with the small number of
events available at present.
The top-right panel is similar but with the sources also
weighted by their X ray flux (both in the computation of the cumulative
fraction of the catalog and also to obtain the data sample simulated
according to the AGN model).
 Although there is still a significant overlap between
the two distributions, the data are in better agreement with the
expectations from the AGN based model and only 0.3\% of the isotropic
simulations have a smaller distance than the data (i.e. the anisotropy
probability defined previously is 99.7\%).
The bottom-left panel is similar to the previous one, i.e. the AGN are also
weighted by their X ray flux, 
but  the cut $|b_G|>15^\circ$ is
imposed, as in \cite{ge08}, what leaves only 19 events in the sample of events
from the Auger Observatory.
This cut in galactic latitude is inspired in the partial incompleteness of
the catalog in regions that could be obscured by the Galaxy.
There are  however  several BAT sources within it (out of 240 AGN seen in the
whole sky, 49 are in the region with $|b_G|<15^\circ$, which corresponds to
a  solid angle of one quarter of the whole sky).
 The qualitative behavior of the
results is very similar as in the previous plot which included the galactic
plane,  being the anisotropy  probability now 98.2\%,
a value comparable to the correlation found in \cite{ge08}.
The bottom-right panel further restricts  the AGN sample to those closer
than 100~Mpc, which are about 40\% of the total AGN sample,
 and in this case $\sim 3.5$\%  of the isotropic simulations
have a value of $ D$ smaller than the data. Compared to
 the simulations
based on the nearby flux-weighted AGN  the distance found
for the data is still typical.

Turning now to the analysis of the HIPASS galaxy catalog studied in
\cite{gh08},  we show in fig.~2 the distribution of distances obtained using
as reference  model the HIPASS objects, weighted by their total HI line flux
$S_{int}$ and subject to different cuts. We not only show the distribution
expected for data simulated according to the particular HIPASS model 
considered in each panel and 
 the isotropic model expectations, 
but also show for comparison the distributions
 obtained for simulated data sets following the SWIFT AGN described before
 (sampling them according to their X ray fluxes, 
without imposing the galactic latitude 
cut and restricting them to those within
100~Mpc). The vertical lines are the corresponding distances $D$ for the 27
highest energy Auger events.

In the top left panel in fig.~2 
we have considered  extragalactic HI sources in the northern 
(NHICAT \cite{wo06}) and  southern  (HICAT
\cite{me04}) HIPASS catalogs. In ref.~\cite {gh08} northern sources 
were cut at $S_{int}>15\ {\rm Jy\ km\ s^{-1}}$ and southern ones
at $S_{int}>7.4\ {\rm Jy\ km\ s^{-1}}$ to have the same completeness level, which was 95\% in both
samples. We think that keeping objects up to different limiting brightnesses 
 introduces however a non-homogeneity in  the two samples 
stronger than what is obtained  keeping the same flux cut for the two samples,
 even if they end up having different completeness levels.  Since the 2DKS
 test just probes the fractional distribution of objects 
 across the sky, it is important to minimize the possible distortions
 introduced in the selection process and hence we will  adopt the flux limit
$S_{int}>7.4\ {\rm Jy\ km\ s^{-1}} $ in both the northern and southern 
subsamples. With this selection one is left 
with 3014 HI galaxies in the field of
view of the Auger Observatory.
As is seen from the plot the distances obtained in the three source models
have a significant overlap, and the value obtained with the 
data set falls just in the overlap region. In particular, being consistent
with the isotropic values the resulting anisotropy probability is not large,
amounting to  44\%.

\begin{figure}[ht]
\centerline{{\epsfig{width=2.in,file=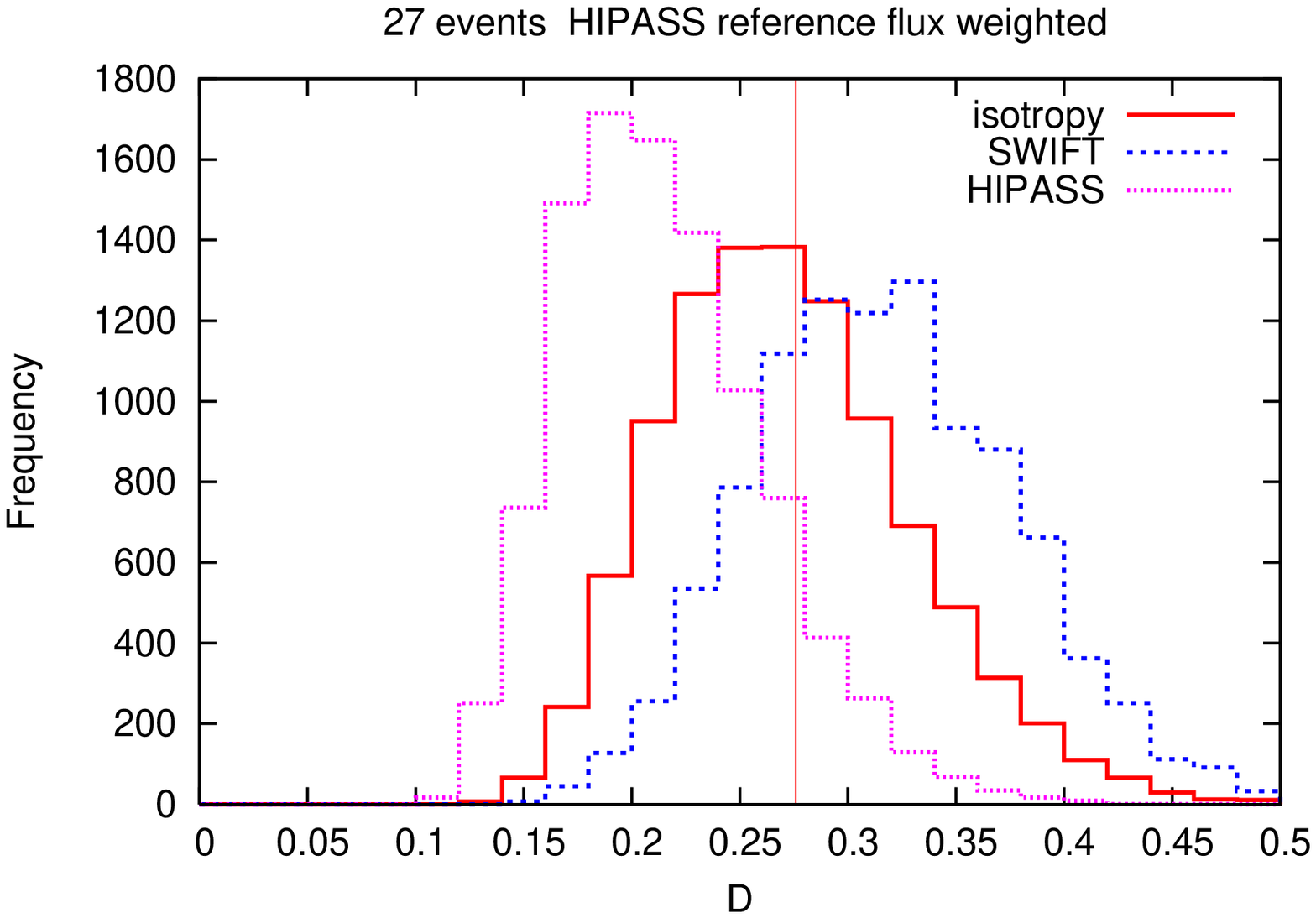,angle=-90}\epsfig{width=2.in,file=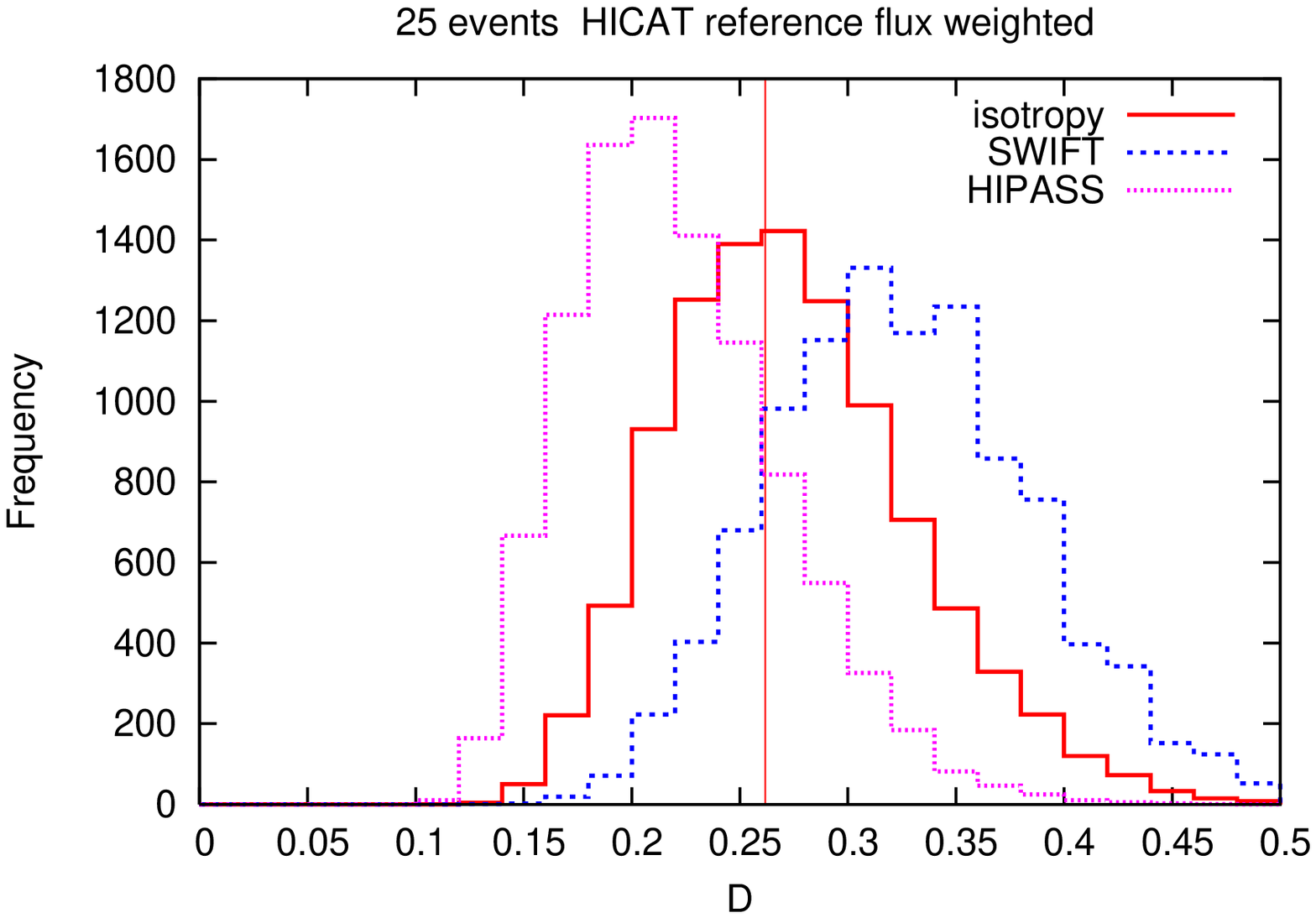,angle=-90}}}
\centerline{{\epsfig{width=2.in,file=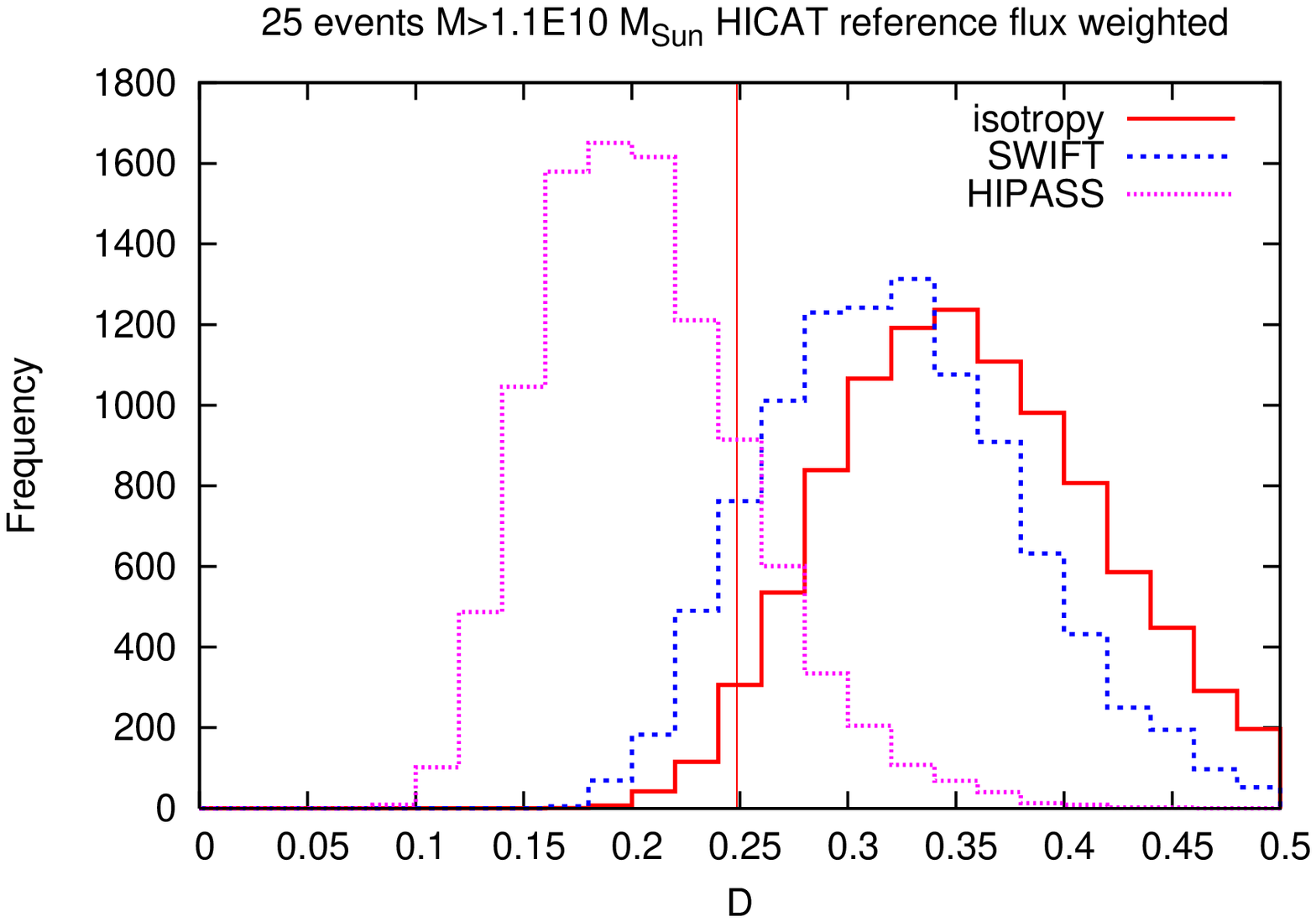,angle=-90}\epsfig{width=2.in,file=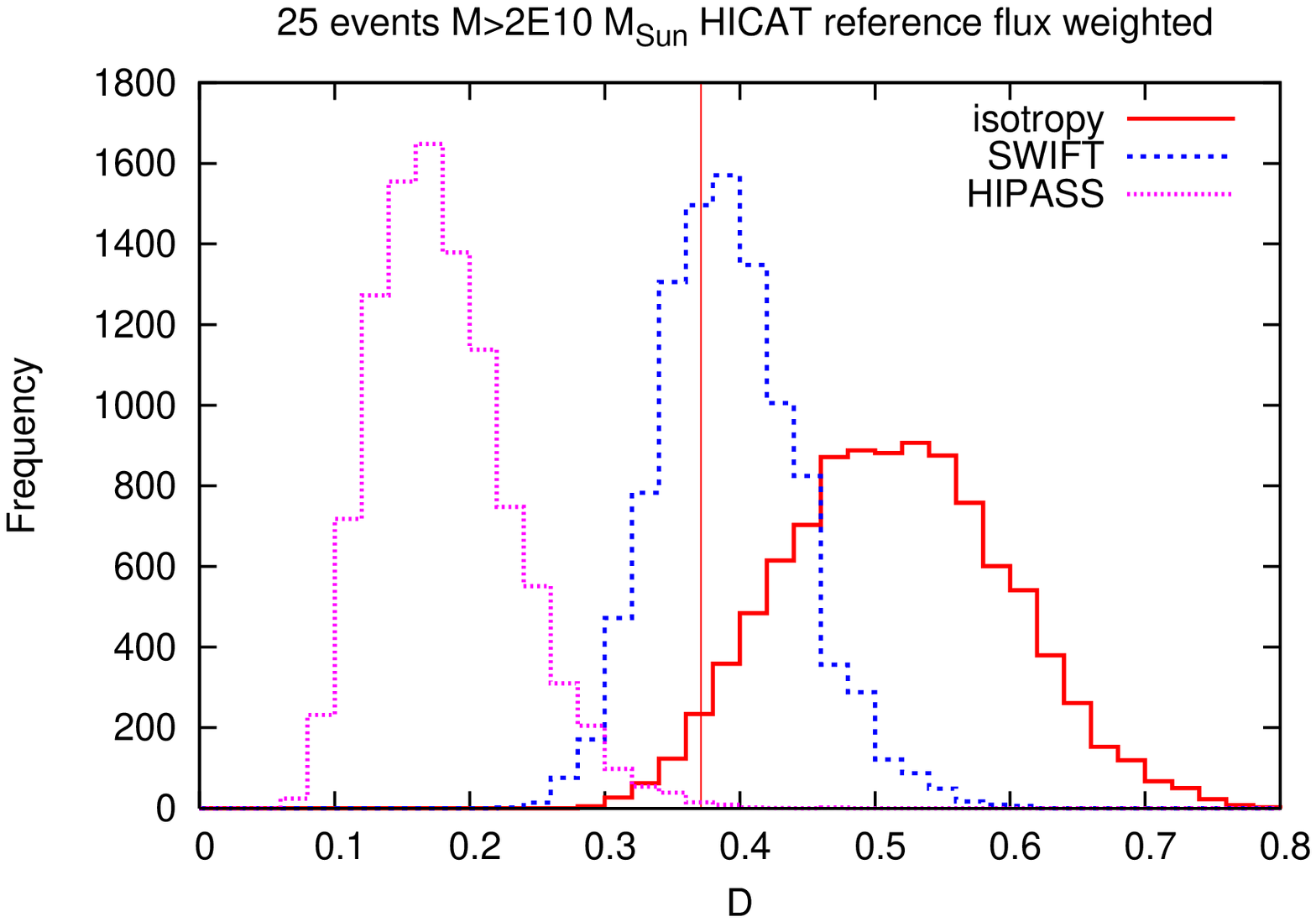,angle=-90}}}
\caption{Distribution of 2DKS distances $D$  between the weighted fractions
  obtained in different HI galaxy models and the fractions in 
simulated data samples
  (according to the same galaxy models (HIPASS), to  AGN models (SWIFT) 
or to isotropy). The vertical lines
  correspond to the values obtained for the Auger data.} 
 \label{fig58.fig}
\end{figure}

The top-right panel in fig.~2
 is similar  but restricted to the HICAT southern sample with declinations
 $\delta<2^\circ$ (and with a
cut $S_{int}>9.4\ {\rm Jy\ km\ s^{-1}}$  as in ref \cite{gh08},
for which the flux limited catalog is 99\% complete, what leaves 1935 galaxies
in this region of the sky).
We also restricted the isotropic and AGN based simulations to the southern
 hemisphere in this plot, and considered only the 25 events from the Auger
 Observatory falling in the same region of the sky.
 The results are
 qualitatively similar to those obtained in the previous plot.
The bottom-left panel 
is further restricted to the galaxies with HI mass\footnote{Following
 \cite{gh08}, we estimate the HI mass content using $M/M_\odot\sim 2.36\times
 10^5 d^2_{Mpc}S_{int}$, with $S_{int}$ expressed in [Jy km s$^{-1}$].} 
$M_{HI}>1.1\times 10^{10}\
M_\odot$ and distances smaller than 100~Mpc, 
for which ref.~\cite{gh08} found a maximal correlation. From
the results depicted it is seen that the data have a distance smaller than
 what is obtained in 
97\% of the isotropic simulations, explaining the large anisotropy
probability found. On the other hand, events simulated according to the same
model (southern radio galaxies with $M_{HI}>1.1\times 10^{10}\
M_\odot$ and $S_{int}>9.4$ Jy km s$^{-1}$) 
have themselves smaller distances (left histogram) than the
isotropic simulations and the distance of the data is quite consistent with
 them.   It is
also interesting to note that simulated data following  the nearby SWIFT AGN 
 have a distribution of distances  (central histogram) somewhat intermediate
 between the HICAT and isotropic ones. For the present number of events  and 
in this
 particular comparison, the data are still quite compatible with both the 
 HICAT and SWIFT models, and is only marginally compatible with isotropy.
The bottom-right panel in fig.~2 is for a reference catalog obtained
 restricting the HICAT sample to the galaxies with HI mass bigger than
 $2\times 10^{10}\ M_\odot$. We see in this case that the anisotropy
 probability is about 97\%, disfavoring the isotropic hypothesis, but
 however the same HICAT model would typically give much smaller distances than
 the data (only 0.3\% of the HICAT simulations have a larger value of $D$ than
 the data).
This implies that the  data from the Auger Observatory 
do not look like a typical realization of
this particular model, 
even if the quoted anisotropy probability is large, showing
 clearly why it is not convenient to call it a correlation probability.
On the other hand, in this test the simulations according to the SWIFT AGN
model considered give a distribution of distances which encompasses quite
 comfortably the value found for the data.
It is however important to keep in mind that the largest distance between the
 AGN based models and the reference HIPASS catalog could be in directions (and
 quadrants) completely different from that giving the largest difference
 between the data and the HIPASS reference catalog, so that the compatibility
 found in this particular test does not imply necessarily that the AGN model
 considered is a good CR source model. In this sense it is important to
 simultaneously check which is the value of the distance between the data and
 the SWIFT AGN reference catalog to better establish the compatibility of the
 data with  this particular source model.

From the different plots in fig.~2 one hence 
concludes that the HIPASS based models considered in the first three
 panels are all consistent with the data results, while the one in the 
fourth panel is not.  It is also worth noting that the
 higher level of rejection of isotropy obtained in the third panel doesn't
 imply necessarily a preference towards sources with
 $M_{HI}>1.1\times 10^{10}M_\odot$.
 
Let us also note that
 in the examples in fig.~2 the data
fractions are compared always to the corresponding fractions in the reference
catalog used (the flux weighted  HIPASS  galaxies with different cuts in each
 panel),  
and the same is done for the  different
source models tested (isotropy, SWIFT AGN and HIPASS ones). In fig.~1, which
 considered the BAT AGN as reference catalog, we
didn't show the corresponding histograms for the HIPASS based models just to
facilitate the presentation, but the same could have clearly been done.
These cross
comparisons are quite useful to discriminate among possible source models,
and we note that the same can be repeated with other reference catalogs and 
source scenarios.

\section{Additional ingredients for the source models}
\subsection{The GZK weight}
\begin{figure}[ht]
\centerline{\epsfig{width=2.in,file=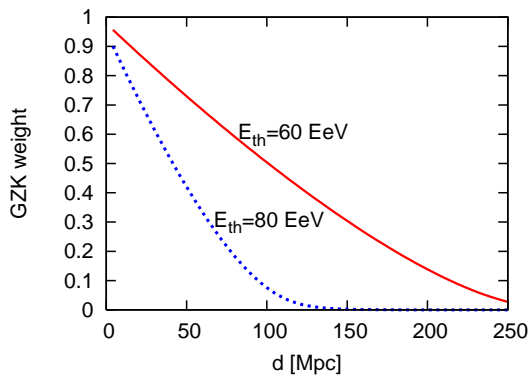,angle=-90}}
\caption{GZK weight vs. distance to the source $d$ for threshold energies
  $E_{th}=60$ and 80~EeV.} 
 \label{fracp.fig}
\end{figure}

A possible additional improvement of the source models would be to weight the
different sources also by a factor accounting for the suppression of
the fluxes above the energy 
threshold considered due to the GZK effect. One can then
account for the expected suppression at the actual distance of each object
instead of just eliminating the sources beyond a given specified distance
(such as in the examples where only sources within 100~Mpc are kept).
 This was indeed considered in the past in alternative analyses
of the CR arrival direction distributions \cite{wax, cuoco}. 
 The required weight factor $W_{GZK}$ is the fraction of the
events produced above a given threshold $E_{th}$
which are able to reach a distance equal to the
distance from the source to us, $d$, with an  energy still above that same
threshold. Assuming a power law spectrum at the sources d$N$/d$E\propto
E^{-s}$ this factor is just
\begin{equation}
W_{GZK}=\frac{s-1}{E_{th}^{-s+1}}\int_{E_i(E_{th},d)}^\infty {\rm d}E\; E^{-s},
\label{frac.eq}
\end{equation}
where $E_i(E_{th},d)$ is the initial energy that a CR must have in order to
survive with $E=E_{th}$ after traveling a distance $d$.
The resulting factors for $E_{th}=60$ and 80~EeV (keep in mind that there are
still significant systematic uncertainties in the energy reconstructed in CR
experiments) are shown in fig.~\ref{fracp.fig},
 computed following reference \cite{horizon}. 
We assumed  a source spectral index of 2.2, but the
results are not much sensitive to the particular value adopted, and considered
 a proton composition.
The suppressions are qualitatively similar for Fe nuclei but are stronger
for intermediate mass nuclei.
Including these factors in the previous comparisons changes the results with
respect to those
obtained using the simple cutoff at 100~Mpc, but not in a drastic way,
although  with larger number of events the proper inclusion of the GZK factors
could become  more important.

\subsection{The faint sources}
Another aspect that may affect the source models is the fact that the
catalogs considered here only contain sources brighter than a certain limit,
and although they may be quite complete above those limiting brightnesses,
the fraction of sources which are  faint
 increases significantly with increasing
distance, and those unobserved sources may actually contribute to the CR
fluxes, possibly giving a more diffuse background. In models where the sources
are weighted by their fluxes the expected 
 contribution from the unobserved sources is however reduced. 
In the past some works \cite{wax,cuoco}
 using the IRAS galaxy catalog, for which the selection effects as a function
 of distance are known, 
corrected for the incompleteness of the catalog by dividing the
 observed density of galaxies at a given distance by the corresponding
 selection function in order to obtain a more 
complete representation of the galaxy distribution. A possible
 drawback of this approach is that one assigns the unobserved galaxies to
 the same locations where bright galaxies are observed, and this
 may not be very
 precise when the galaxies are sparsely sampled, as happens at large
 distances. For smaller catalogs, such as the SWIFT one, even if a selection
 function were known it would not be very realistic to assume that
 the unobserved 
faint AGN are in the same locations as the bright ones. 
Anyway, if one restricts the sources to the nearby ones 
(using a distance cutoff or weighting them by the GZK attenuation factors),
in the models where the sources are weighted
 by their fluxes the contribution from the faint
unobserved sources is not large, so that the model expectations obtained 
should still be reasonably accurate.
For instance, if one models the luminosity distribution of the sources
according to a Schechter function\footnote{This discussion can be easily
  extended to the case of a broken power law luminosity function, as that
  used to fit SWIFT data \cite{bat}.}, d$N/$d$L\propto L^a\exp(-L/L_*)$, where a
typical value for the faint end slope is $a\simeq -1$, one finds that the
fraction $F(z)$ of the flux observed above a given flux limit $f_{min}$ from the
sources at redshift $z$ is $F(z)\simeq \exp(-f_{min}/f_*)$, where $f_*$ is the
flux received from an $L_*$ galaxy lying at redshift $z$. 
 In the scenarios in which one assumes
that the cosmic ray luminosities are proportional to the observed catalog
luminosities (flux weighted scenarios)  the fraction $\eta$
of the total  cosmic ray flux which is accounted by the model will then be
\begin{equation}
\eta={{\int {\rm d}z\ F(z) W_{GZK}(z)}\over {\int {\rm d}z\  W_{GZK}(z)}}.
\end{equation}
Defining the
characteristic depth of the survey $z_*$ as that corresponding to the redshift
at which $f_*=f_{min}$, one finds for instance that for the representative
value $z_*=0.3$ (i.e. about 130~Mpc), the fraction $\eta$ turns out to be 
88\% if one uses the GZK attenuation factor corresponding to $E_{th}=80$~EeV,
while it is 66\% if one adopts $E_{th}=60$~EeV. For $z_*=0.05$ (corresponding
to a survey depth of about 200~Mpc), the corresponding fractions are 95\% and
82\% respectively. Moreover, these fractions become larger in models in which
the cosmic ray sources are restricted  to those with absolute luminosities
above a certain threshold, as was the case for the HIPASS based models
restricted to galaxies more massive than a certain limiting mass.

\section{Prospects with increased statistics}
Let us now illustrate the possible discrimination power of the 2DKS test with
future increased data samples. For definiteness we 
use in these examples the GZK factor corresponding to $E_{th}=80$~EeV, for
which sources beyond 100~Mpc have essentially a negligible contribution and
closer ones are weighted non-trivially, 
and weight the SWIFT and HIPASS sources by their respective fluxes. We
consider the HIPASS galaxies (both the northern and southern ones) 
with $S_{int}>7.4\ {\rm Jy\ km\ s^{-1}}$
 and $M_{HI}>10^{10}\ M_\odot$. 
The left panels in fig.~4 show the distribution of $D$ values expected for 
$n=50$ and
100 events using as reference catalog the SWIFT one, and the right panels
 show the results using the HIPASS reference catalog. 
As a general trend, one can see that the $D$ distribution of the data sampled
according to  the 
reference catalog considered just scales as $n^{-1/2}$, peaking
at about $D_{peak}\simeq 0.2\sqrt{25/n}$, 
being quite independent from the particular catalog adopted
 (see also figs. 1 and 2). On the other hand, for the scenarios differing from
the reference one (either one based on a different catalog or the
isotropic hypothesis), the distribution of $D$ values tends to a given non-zero
average value, with the associated dispersion decreasing as $n^{-1/2}$, as is
also seen in fig.~4.

\begin{figure}[ht]
\centerline{{\epsfig{width=2.in,file=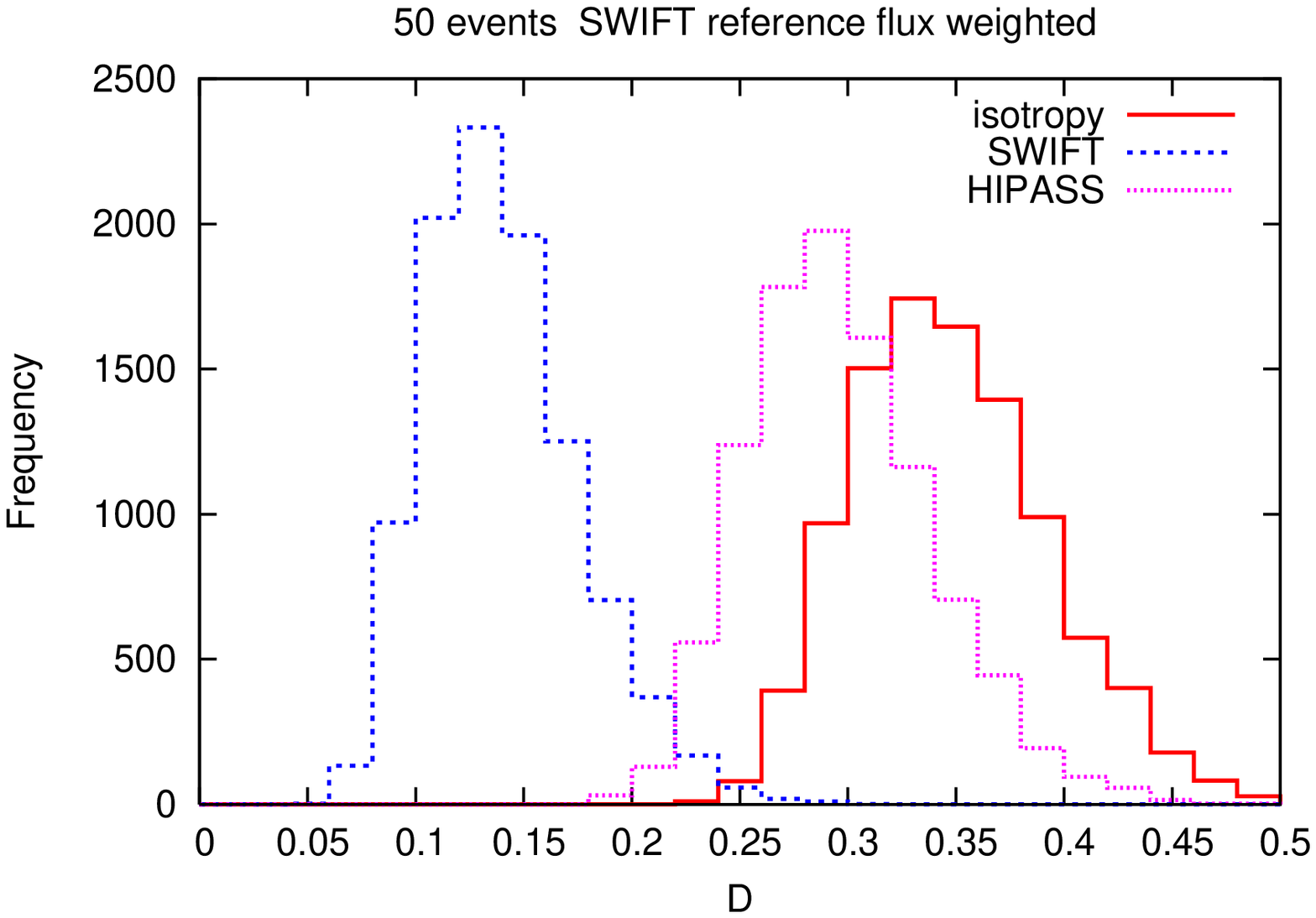,angle=-90}\epsfig{width=2.in,file=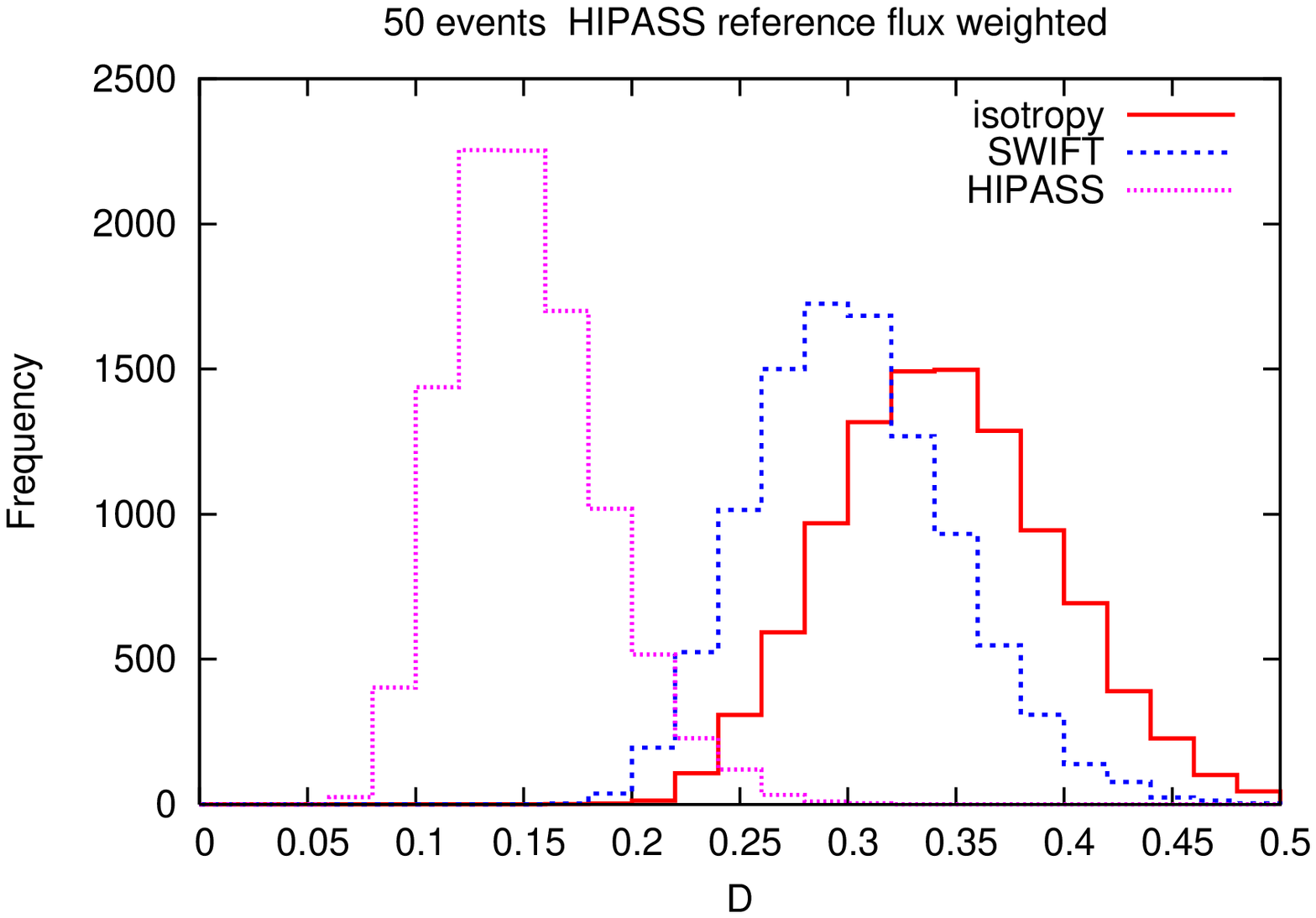,angle=-90}}}
\centerline{{\epsfig{width=2.in,file=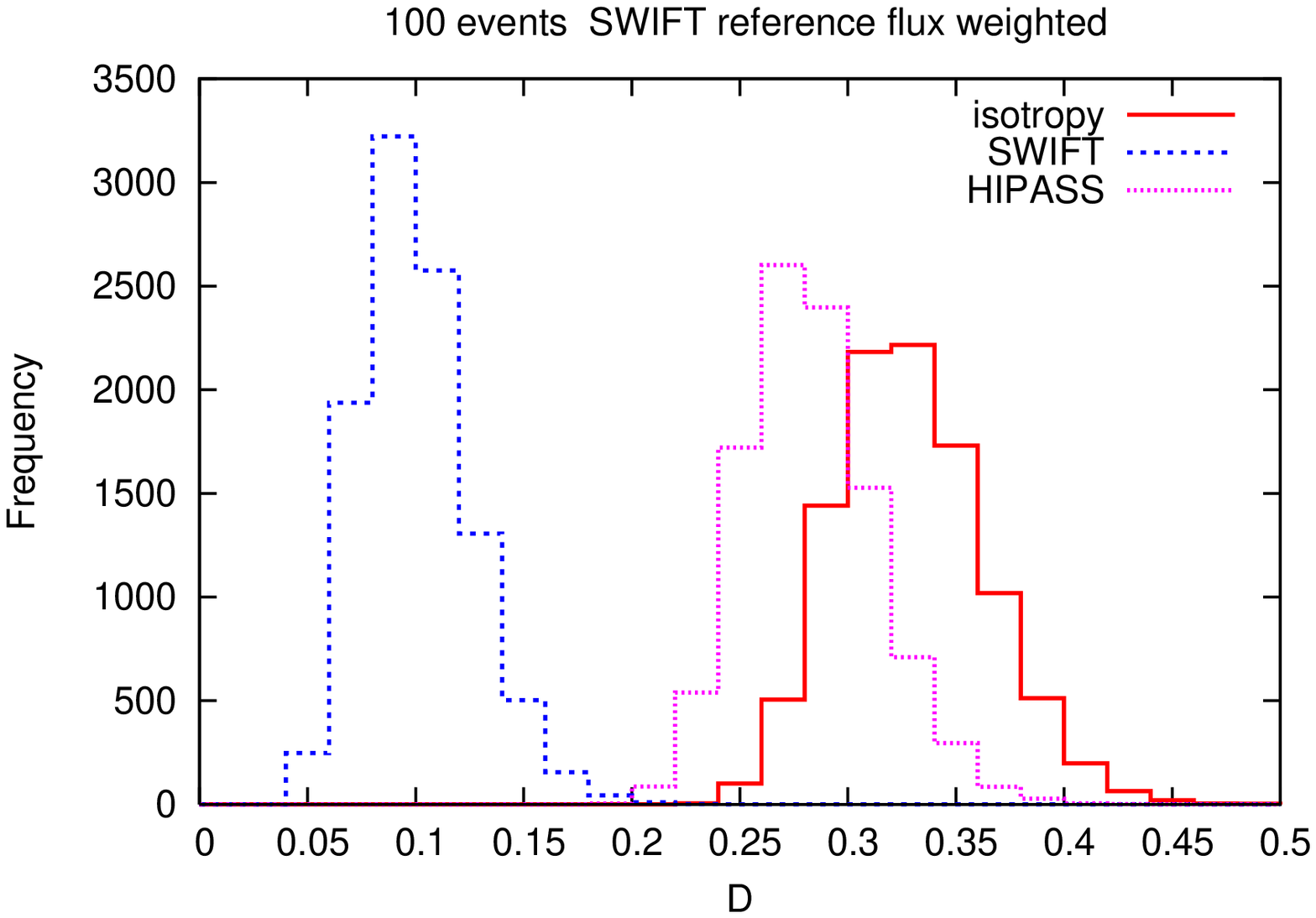,angle=-90}\epsfig{width=2.in,file=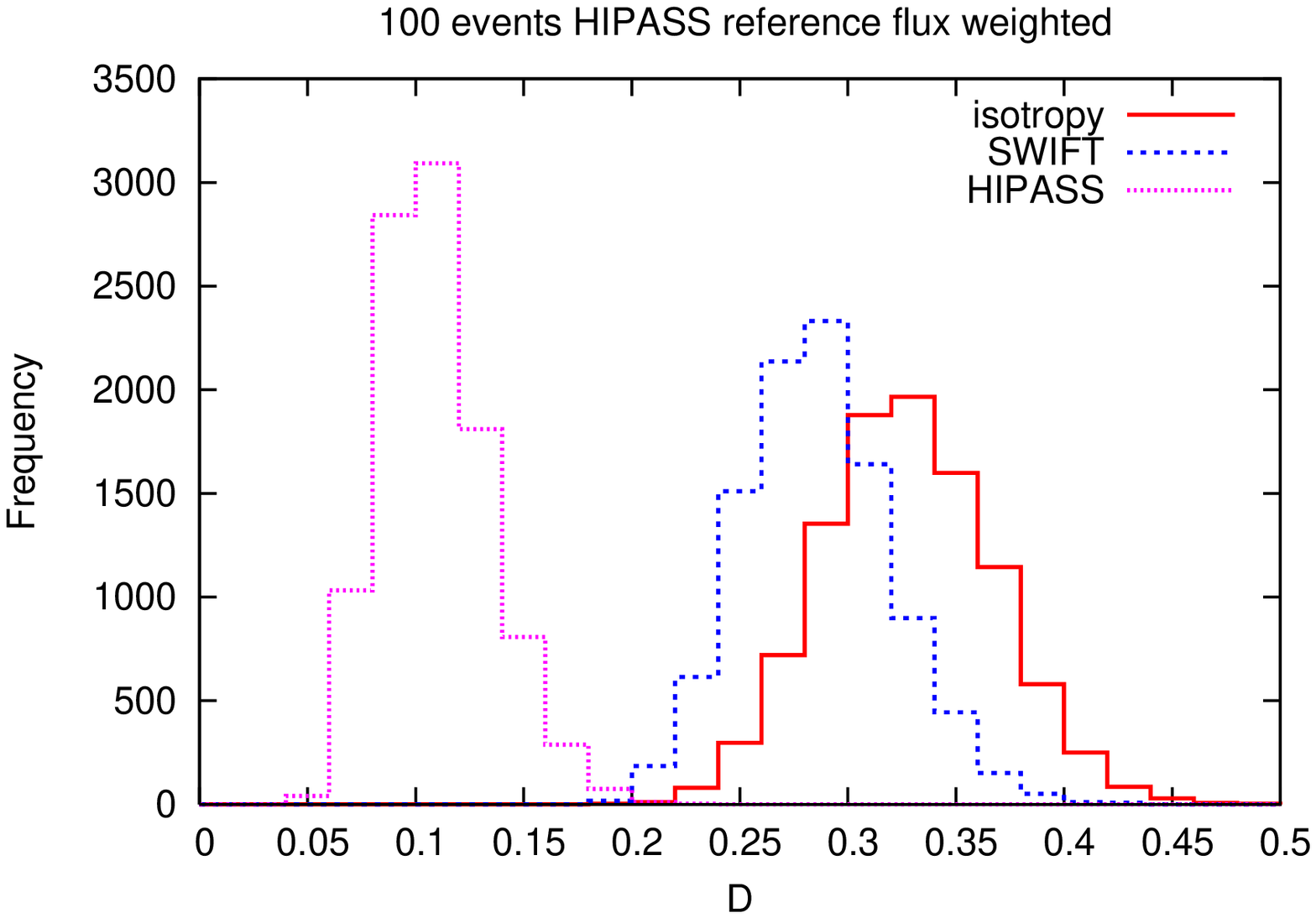,angle=-90}}}
\caption{Distribution of 2DKS distances $D$  between the weighted fractions
  obtained in models based on AGN from SWIFT,  galaxies from HIPASS and
  isotropy, using as reference catalog
 the AGN one (left) or HIPASS one (right),
  for data samples of 50 events (top) or 100 events (bottom). } 
 \label{fig4.fig}
\end{figure}

It is clear that for 100 events 
 the distributions of the different models become quite
separated from each other, and hence this method will have the ability to
confidently tell apart these scenarios after a few years of operation of the
Auger Observatory. However, as was apparent from figs.~1 and 2, according to
the Kolmogorov-Smirnov test  the Auger Observatory 
data analyzed are still consistent with many different possible
scenarios (except isotropy to a certain level). One has to keep in mind
however that the energy threshold for this 
dataset was selected by the Auger Collaboration  as that
maximizing the correlation with AGN from the VC catalog, and hence although
the discrepancy with isotropy in some of the tests is reassuring, it does not
provide a totally independent test of anisotropy.

Finally, it is also useful to consider how would the different sky
distributions look for a large number of simulated events, in which case the
effects of sampling 
fluctuations tend to vanish. Identifying the quadrant responsible
for the largest fractional difference $D$ will hence 
characterize the most important
feature discriminating among alternative
scenarios in these tests.
In particular, when comparing  the isotropic and AGN based distributions
we find that the largest value of $D$ arises, in the large data size limit,
most frequently from the fourth quadrant, measured in a counter clockwise
order starting from the top-right one, with respect to the direction
$(146^\circ, -14^\circ)$, for
which 66\% of the weighted AGN lie while only 36\% of the isotropic events
are found.
On the other hand for the HIPASS (north and south with $S_{int}>7.4\ {\rm Jy\ km\ s^{-1}}$) 
 flux weighted catalog and
the isotropic simulations one finds that 
the biggest difference is obtained in 
the first quadrant with respect to the approximate direction
$(225^\circ,-55^\circ)$, 
for which the HIPASS simulations have a fraction of about 14\% while the
isotropic ones of 30\%. Also in the fourth quadrant with respect to the 
 direction
$(140^\circ,-30^\circ)$ there is a similar difference,  
with the HIPASS simulations having a fraction of about 45\% while the
isotropic ones of 30\%. 
It is mainly these characteristics  that
may hence allow to differentiate among the possible source
 populations using this method once a larger number of events become available.

\section{Discussion}
We have considered in some detail the application of the 2DKS test to the
study of the distribution of the arrival directions of the UHECRs, extending
recent studies of this kind and discussing possible generalizations of the
method, in particular introducing the cross comparisons between the
distributions of the distances  obtained in the different scenarios (not just
the isotropic one) and the value of $D$ obtained with the data. 
It is important to keep in mind that what the 2DKS test  probes is whether the
overall distribution of the data is approximately proportional to that of the
model considered. This makes the method quite sensitive to the cuts and weights
adopted since if only a subsample of the objects in the catalog are the
actual sources their distribution in the sky 
may be quite different from the overall
one. In addition, if the assumption that the actual UHECR source luminosities
are proportional to the luminosities measured in some particular wavelength
(as was the case in most of the models considered) does not hold, the overall
CR distribution can be quite different from the one resulting from
 the model, even if the sources involved 
are indeed the true CR sources. 
This may be the strongest limitation of the 2DKS
method, and may be a particularly delicate issue if the number of sources
contributing to the UHECR events above the threshold energy considered is not
large. 
In this sense the test of correlations performed by the Auger Collaboration
is of a
different nature, because it just looked for the existence of some AGN
within a certain distance and at less than a given angle from the events,
without requiring an overall distribution in the sky proportional to the 
number of objects in the catalog 
(eventually weighted in some specific way). 

As we have shown here, with the enlarged data samples  expected in the next
few years with the continuous operation of the 
Auger Observatory it will become possible to test in a
significant way several possible models for the distribution of the UHECRs. If
the actual CR distribution is different from the model assumptions, this can
be put in evidence by comparing the distribution of the simulations
according to the particular model and the actual data.
The distribution of distances for a model, when one uses the same model as
reference, peaks at $D_{peak}\simeq 1/\sqrt{n}$, while on the other hand
models different from the reference one tend to give rise to 
larger average distances. Regarding the isotropic simulations, the more
anisotropic is the reference catalog
 to which they are compared, the larger will
be the distances obtained, as is apparent from figs. 1 and 2.
The 2DKS method can certainly be used as a way to disproof isotropy, and the
most efficient way to achieve  this will be when using as reference 
scenario the one
closest to the actual source distribution.

\section*{Acknowledgments}

We are grateful to Jack Tueller for providing us the 22 months BAT catalog
 prior to publication and to  I. Wong for sending us the NHICAT catalog. 
This work is supported by
 ANPCyT (grant PICT 13562-03) and CONICET (grant PIP 5231). We want to thank
 Paul Sommers for discussions.

\end{document}